\documentclass[twocolumn,english,showpacs,preprintnumbers,jap]{revtex4-1}
\usepackage[latin9]{inputenc}
\usepackage{color}
\usepackage{amsmath}
\usepackage{amssymb}
\usepackage{graphicx}
\usepackage{esint}

\makeatletter
\@ifundefined{textcolor}{}
{%
 \definecolor{BLACK}{gray}{0}
 \definecolor{WHITE}{gray}{1}
 \definecolor{RED}{rgb}{1,0,0}
 \definecolor{GREEN}{rgb}{0,1,0}
 \definecolor{BLUE}{rgb}{0,0,1}
 \definecolor{CYAN}{cmyk}{1,0,0,0}
 \definecolor{MAGENTA}{cmyk}{0,1,0,0}
 \definecolor{YELLOW}{cmyk}{0,0,1,0}
}

\makeatother

\usepackage{babel}
\begin{document}

\title{Fano interference and a slight fluctuation of the Majorana hallmark}

\author{A. C. Seridonio$^{1,2}$, E. C. Siqueira$^{2}$, F. A. Dessotti$^{2}$,
R. S. Machado$^{2}$, and M. Yoshida$^{1}$}

\affiliation{$^{1}$Instituto de Geociências e Ciências Exatas - IGCE, Universidade
Estadual Paulista, Departamento de F\'{i}sica, 13506-970, Rio Claro,
São Paulo, Brazil\\
 $^{2}$Departamento de F\'{i}sica e Qu\'{i}mica, Universidade Estadual
Paulista, 15385-000, Ilha Solteira, São Paulo, Brazil}
\begin{abstract}
According to the Phys. Rev. B 84, 201308(R) (2011), an isolated Majorana
state bound to one edge of a long enough Kitaev chain in the topological
phase and connected to a quantum dot, results in a robust transmittance
of $1/2$ at zero-bias. In this work, we show that the removal of
such a hallmark can be achieved by using a metallic surface hosting
two adatoms in a scenario where there is a lack of symmetry in the
Fano effect, which is feasible by coupling the Kitaev chain to one
of these adatoms. Thus in order to detect this feature experimentally,
one should apply the following two-stage procedure: (i) first, attached
to the adatoms, one has to lock AFM tips in opposite gate voltages
(symmetric detuning of the levels $\Delta\varepsilon$) and measure
by an STM tip, the zero-bias conductance; (ii) thereafter, the measurement
of the conductance is repeated with the gates swapped. For $\left|\Delta\varepsilon\right|$
away from the Fermi energy and in the case of strong coupling tip-host,
this approach reveals in the transmittance, a persistent dip placed
at zero-bias and immune to the aforementioned permutation, but characterized
by an amplitude that fluctuates slightly around $1/2$. However, in
the case of a tip acting as a probe, the adatom decoupled from the
Kitaev chain becomes completely inert and no fluctuation is observed.
Therefore, the STM tip must be considered in the same footing as the
``host+adatoms'' system. As a result, we have found that despite
the small difference between these two Majorana dips, the zero-bias
transmittance as a function of the symmetric detuning yields two distinct
behaviors, in which one of them is unpredictable by the standard Fano's
theory. Therefore, to access such a non trivial pattern of Fano interference,
the hypothesis of the STM tip acting as a probe should be discarded.
\end{abstract}


\maketitle

\section{Introduction}

\label{sec:intro}

Majorana fermions are particles that constitute their own antiparticles.
Such a proposal was made almost a century ago by Ettore Majorana in
the context of high-energy physics. In solid state systems, these
exotic particles are not fundamental but emerge as quasiparticle excitations
\cite{key-24}. This species of excitation is ranked as non-Abelian
anyons and obeys an unusual quantum statistics. Its most remarkable
property lies on the possibility of bounding two far apart Majoranas
that define an unique nonlocal Dirac fermion. Once this spatially
delocalized state is occupied, it yields a robust qubit decoupled
from the surroundings, thus avoiding decoherence due to perturbations.
This protected qubit then enlarges the feasibility to make these blocks
as essential to the accomplishment of a topological quantum computer.
Thus in the last few years the quest for devices nesting Majorana
fermions has received much attention from the community of researchers
working with quantum computing \cite{key-33,key-34,key-35,key-36,key-37}.

To the best knowledge, the superconductor state is considered suitable
for the emergence of Majorana excitations. Superconductivity lies
on Cooper-pair condensation and spontaneous breaking of charge conservation,
thus leading to the superposition of electrons and holes. However,
s-wave superconductivity arises from electrons with opposite spins
that results in distinct operators for creation and annihilation of
quasiparticles, thus preventing the realization of Majorana bound
states (MBSs). To support them, a spinless superconductor is indeed
required. Such conditions can be found in the topological phase of
the Kitaev chain \cite{key-333}, which offers the proper environment
to sustain Majoranas. The Majoranas are zero-energy modes, in particular,
placed at the edges of this chain.

\begin{figure}
\includegraphics[width=0.52\textwidth,height=0.26\textheight]{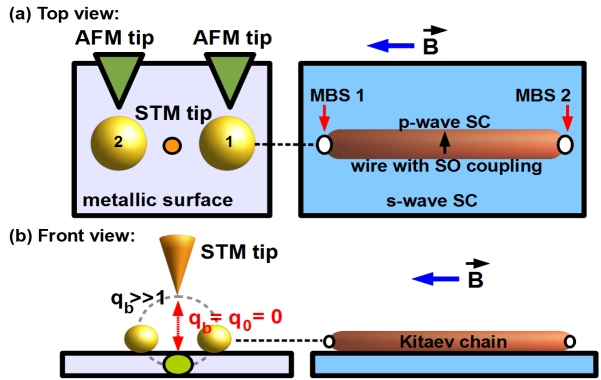}
\caption{\label{Fig1}\textcolor{black}{(Color online) Two perspectives of the same apparatus:
in panel (a) we have the top view, while (b) represents the front view. In both,
Majorana bound states (MBSs) appear lying on a long enough Kitaev chain
within the topological phase [right side of panels (a) and (b)], which can be accomplished as proposed
experimentally in Ref. {[}\onlinecite{key-301}{]}: s-wave superconductivity (SC) inducing p-wave pairing in
a semiconducting wire with strong spin-orbit interaction (SO) and crossed by a perpendicular magnetic field $\vec{B}$.
Here we follow such a proposal by adding an STM tip nearby a metallic surface
coupled to two adatoms, in which one of them is hybridized with a
MBS 1 (a half-electron state). The MBS 1 is connected far apart to
a MBS 2. The AFM tips are employed to tune the levels of the adatoms. This device explores the lack of symmetry in Fano interference,
which is detectable via the zero-bias conductance. The parameters
$q_{0}$ and $q_{b}$ are Fano factors due to the interference between
the different paths taken by the electrons from the tip to the surface.
When $q_{b}\gg1$ the hybridization between the tip and the adatoms
is stronger than the hybridization to the surface. In this case the
electrons tunnel to the surface throughout the adatoms. In contrast,
for $q_{0}=0$ the electrons tunnel directly to the surface. The green-circle
represents the site of the host side-coupled to the adatoms.}}
\end{figure}

The engineering of a sample with \emph{p}-wave superconductivity can
be achieved experimentally by proximity effect. It is known that a
\emph{s}-wave superconductor nearby a semiconducting nanowire with
strong spin-orbit interaction and crossed by a magnetic field, induces
\emph{p}-wave superconductivity on the latter system \cite{key-371,key-372,key-373,key-374,key-375,key-376,key-377,key-378,key-379}.
Additionally, the existence of Majoranas are predicted in the fractional
quantum Hall state with filling factor $\nu=5/2$ \cite{key-311},
in three-dimensional topological insulators \cite{key-31} and at
the core of superconducting vortices \cite{key-28,key-29,key-30}.
In this scenario, quantum transport becomes a sensible tool for detecting
Majorana quasiparticles. Particularly in Ref. {[}\onlinecite{key-25}{]},
it was predicted for the experimental setup of a single quantum dot
(QD) side-coupled to a Majorana state, that the zero-bias peak (ZBP)
for the conductance should be given by the robust Majorana hallmark
$\mathcal{G}=0.5\mathcal{G}_{0}$, where $\mathcal{G}_{0}=e^{2}/h$
is the background conductance. We highlight that in Ref. {[}\onlinecite{key-27}{]},
E. Vernek\textit{ et al. }have determined that such an amplitude arises
from the leaking of the Majorana state into the QD.

Experimentally, a persistent ZBP has been observed in transport measurements
through a setup composed by a nanowire of indium antimonide merged
to gold and niobium titanium nitride \cite{key-301}. In this aforementioned
system, Majoranas are supposed to exist due to the ZBP that stands
up to a wide range of magnetic fields and gate voltages. Such a robustness
of the ZBP has also been found in the analogous system of a superconductor
of aluminium close to a nanowire of indium arsenide \cite{key-302}.
Moreover, we stress that the ZBP feature may also have another physical
origin, for instance, the Kondo effect \cite{Hewson,QD1,QD2,STM12,STM5,STM6,AHCNeto,STM13}.

In this context, an apparatus based on Fano effect \cite{Fano1,Fano2}
becomes an alternative approach to detect a Majorana state. Here we
benefit of this mechanism, an interference phenomenon found in systems
where tunneling channels compete for the electron transport. This
effect can be detectable by the Scanning Tunneling Microscope (STM),
a device made by a metallic tip that detects, for low enough temperatures,
the transmittance through a system by measuring the differential conductance
\cite{STM12,STM5,STM6,AHCNeto,STM13}. Thus we have studied theoretically
the conductance probed by an STM tip of a metallic surface coupled
to two adatoms, in which one of them is coupled to a MBS hosted by
a long enough Kitaev chain in the topological phase.
We should remark that nowadays such a chain is achievable experimentally
as found in Ref. {[}\onlinecite{key-301}{]}, whose system becomes the most promising candidate to our proposal
{[}see Fig. \ref{Fig1}{]}.

Additionally, we have considered in the model two Atomic Force Microscope
(AFM) tips capacitively coupled to the adatoms, just in order to tune
their levels as proposed in Ref. {[}\onlinecite{key-3}{]}. Our
approach employs the spinless Hamiltonian of Ref. {[}\onlinecite{key-25}{]}
in combination with the equation-of-motion procedure for the Green's
functions.

By determining the transmittance of this setup, we have found Fano
profiles due to the coupling between the setup of the adatoms and
an isolated MBS. For the setup decoupled from this MBS, the direct
and the mixed Green's functions are symmetric with respect to the
labels $1$ and $2$ that designate the parameters of the adatoms.
In the opposite limit, this symmetry property is broken and the swap
of the indexes $1\leftrightarrow2$ leads to a lack of symmetry in
the Fano profile.

\textit{This lack of symmetry can be accessed experimentally by performing
the following proposed two-stage procedure: (i) first, attached to
the adatoms, one has to lock AFM tips in opposite gate voltages (symmetric
detuning of the levels }$\Delta\varepsilon$\textit{) and measure
by an STM tip, the zero-bias conductance; (ii) thereafter, the measurement
of the conductance is repeated with the gates swapped.}

As a result of this method and the Fano regime as well, the transmittance
for $\left|\Delta\varepsilon\right|$ away from the Fermi energy exhibits
a zero-bias dip persistent against the permutation of the gate voltages.
For the case in which the STM acts as a probe of the LDOS (local density
of states) for the ``host+adatoms'' system, the adatom decoupled
from the Kitaev chain plays no role and the typical Majorana hallmark
is verified: a robust zero-bias transmittance characterized by an
amplitude of $1/2$ as that found in Ref. {[}\onlinecite{key-25}{]}
for a single QD setup. On the other hand, for the STM in the same
footing as the ``host+adatoms'' system, a slight fluctuation around
the amplitude of $1/2$ manifests as a straight aftermath of the two-stage
procedure in combination with the adatom free of the MBS. However,
despite the small difference between these two Majorana dips, each
one leads to a particular Fano lineshape for the zero-bias transmittance
as a function of the symmetric detuning. Therefore, we demonstrate
in this work that the assumption of the STM as a probe tip is not
enough to reveal the unexpected pattern of Fano interference for the
proposed setup of Fig. \ref{Fig1}.

This paper is organized as follows. In Sec. \ref{sec2}, we show the
theoretical model for the system sketched in Fig. \ref{Fig1} as well
as the derivation of the transmittance. The Green's functions of the
adatoms are also presented in this section. The results appear in
Sec. \ref{sec3} and in Sec. \ref{sec4}, we summarize the conclusions.

\section{Theoretical Model}

\label{sec2}

\subsection{Hamiltonian}

\label{sec2A}

The system we investigate is described according to the Hamiltonian

\begin{equation}
\mathcal{H}_{\text{{total}}}=\mathcal{H}_{\text{host+ads}}+\mathcal{H}_{\text{tip}}+\mathcal{H}_{\text{tun}}.\label{eq:Total}
\end{equation}

In order to mimic the system outlined in Fig. \ref{Fig1}, we follow
the spinless Hamiltonian proposed by Liu \textit{et al.} \cite{key-25},
taking two adatoms into account, which reads
\begin{align}
\mathcal{H}_{\text{host+ads}} & =\sum_{k}(\varepsilon_{k}-\mu_{\text{host}})c_{k}^{\dagger}c_{k}+\sum_{j}\varepsilon_{j}d_{j}^{\dagger}d_{j}\nonumber \\
 & +V(\sum_{jk}c_{k}^{\dagger}d_{j}+\text{{H.c.}})+i\epsilon_{M}\eta_{1}\eta_{2}\nonumber \\
 & +\lambda(d_{1}-d_{1}^{\dagger})\eta_{1},\label{eq:TIAM}
\end{align}
where the electrons in the host are described by the operator $c_{k}^{\dagger}$
($c_{k}$) for the creation (annihilation) of an electron in a quantum
state labeled by the wave number $k$, energy $\varepsilon_{k}$ and
chemical potential $\mu_{\text{host}}$. For the adatoms, $d_{j}^{\dagger}$
($d_{j}$) creates (annihilates) an electron in the state $\varepsilon_{j}$,
with $j=1,2$. $V$ is the hybridization of the adatoms with the host.
In particular for $j=1$, the adatom 1 is coupled to the MBS 1 described
by the operator $\eta_{1}^{\dagger}=\eta_{1}$. The strength of this
coupling is $\lambda$. The MBS 2 given by $\eta_{2}^{\dagger}=\eta_{2}$
is connected to the MBS 1 via the coefficient $\epsilon_{M}\sim e^{-L/\xi}$,
with $L$ being the distance between the MBSs and $\xi$ the coherence
length. It is worth mentioning that the present spinless model supposes
a strong magnetic field over the entire system, which leads to a large
Zeeman splitting where the higher levels are not energetic favorable
at low temperatures. In this situation, one spin component plays no
role and the spin degrees of freedom can be ignored.

The second part of Eq. (\ref{eq:Total}) is described by the Hamiltonian

\begin{equation}
\mathcal{H}_{\text{tip}}=\sum_{q}(\varepsilon_{q}-\mu_{\text{tip}})b_{q}^{\dagger}b_{q},\label{eq:STM}
\end{equation}
which corresponds to free electrons ruled by fermionic operators $b_{q}^{\dagger}$
and $b_{q}$ in the STM tip, with energy $\varepsilon_{q}$ and chemical
potential $\mu_{\text{tip}}$.

To perform the coupling between Eqs. (\ref{eq:TIAM}) and (\ref{eq:STM}),
we have to define the tunneling Hamiltonian

\begin{eqnarray}
\mathcal{H}_{\text{tun}} & = & w(f_{t}^{\dagger}\psi_{0}+\text{{H.c.})},\label{eq:Tun}
\end{eqnarray}
where $w$ is the STM tip-host coupling,

\begin{equation}
f_{t}=\sum_{q}b_{q}\label{eq:ft}
\end{equation}
is for the edge of the STM tip,

\begin{equation}
\psi_{0}=f_{0}+(\pi\Gamma\rho_{0})^{1/2}q_{0}\sum_{j}d_{j}\label{eq:PSI_R-1-1}
\end{equation}
is the field operator that accounts for Fano interference,
\begin{equation}
f_{0}=\sum_{k}c_{k}\label{eq:f0}
\end{equation}
represents the host site laterally coupled to the adatoms {[}see the
green-circle of the host outlined in Fig. \ref{Fig1}{]},

\begin{equation}
\Gamma=\pi V^{2}\rho_{0}\label{eq:AP}
\end{equation}
is the Anderson parameter, with $\rho_{0}=\frac{1}{2D}$ as the density
of states for the surface without adatoms, $D$ is the band half-width
and
\begin{equation}
q_{0}=(\pi\Gamma\rho_{0})^{-1/2}\left(\frac{\tilde{V}}{w}\right)\label{eq:Fano_q}
\end{equation}
is the Fano factor of interference \cite{FanoSTM}, with $\tilde{V}$
as the couplings between the STM tip and the adatoms. Notice that
due to Eqs. (\ref{eq:PSI_R-1-1}) and (\ref{eq:Fano_q}), the limit
$q_{0}\gg1$ represents the situation in which the tip is highly hybridized
with the adatoms, while in the opposite regime $q_{0}=0$, the tip
is strongly connected to the surface {[}see Fig. \ref{Fig1}{]}. As
the former case in presence of a MBS still obeys the standard Fano's
theory, in this work we will focus on the latter, where we can find
a non trivial Fano interference. Such a point will be discussed in
Sec. \ref{sec3}.

\subsection{Calculation of the transmittance }

\label{sub:2b}

\subsubsection{The STM tip as a probe}

\label{sub:2b2}

By applying the linear response theory, in which the STM tip is considered
as a probe, it is possible to show that the zero-bias conductance
is given by
\begin{equation}
\mathcal{G}(0)=\frac{e^{2}}{h}(2\pi w)^{2}\int\rho_{\text{LDOS}}(\varepsilon)\rho_{\text{tip}}(\varepsilon)\left(-\frac{\partial f_{F}}{\partial\varepsilon}\right)d\varepsilon,\label{eq:DC}
\end{equation}
where $e$ is the electron charge, $h$ is the Planck constant, $\rho_{\text{LDOS}}(\varepsilon)$
is the LDOS of the ``host+adatoms'' system, $\rho_{\text{tip}}(\varepsilon)$
as the DOS of the STM tip and $f_{F}$ is the Fermi-Dirac distribution.
The total transmittance is then defined as follows:

\begin{equation}
\mathcal{T}_{\text{{probe}}}(\varepsilon)=(2\pi w)^{2}\rho_{\text{LDOS}}(\varepsilon)\rho_{\text{tip}}(\varepsilon).\label{eq:transmit}
\end{equation}

To obtain the LDOS, we follow Ref. {[}\onlinecite{key-3}{]} by
introducing the retarded Green's function

\begin{align}
\mathcal{R}_{\psi_{0}\psi_{0}} & =-\frac{i}{\hbar}\theta\left(t\right){\tt Tr}\{\varrho_{\text{{host+ads}}}[\psi_{0}\left(t\right),\psi_{0}^{\dagger}\left(0\right)]_{+}\}\label{eq:PSI_R}
\end{align}
for the field operator of Eq. (\ref{eq:PSI_R-1-1}) in the time domain
$t$, where $\theta\left(t\right)$ is the Heaviside function, $\varrho_{\text{{host+ads}}}$
is the density matrix of the system described by the Hamiltonian in
Eq. (\ref{eq:TIAM}) and $[\cdots,\cdots]_{+}$ is the anticommutator
of Eq. (\ref{eq:PSI_R-1-1}) at distinct times. From Eq.~(\ref{eq:PSI_R}),
the LDOS of the host can be obtained as
\begin{equation}
\rho_{\text{LDOS}}(\varepsilon)=-\frac{1}{\pi}{\tt Im}(\tilde{\mathcal{R}}_{\psi_{0}\psi_{0}}),\label{eq:FM_LDOS}
\end{equation}
where $\tilde{\mathcal{R}}_{\psi_{0}\psi_{0}}$ is the Fourier transform
of $\mathcal{R}_{\psi_{0}\psi_{0}}$ in the energy domain $\varepsilon$.
Analogously, we have

\begin{equation}
\rho_{\text{tip}}(\varepsilon)=-\frac{1}{\pi}{\tt Im}(\tilde{\mathcal{R}}_{f_{t}f_{t}}),\label{eq:STM_tip}
\end{equation}
with

\begin{align}
\mathcal{R}_{f_{t}f_{t}} & =-\frac{i}{\hbar}\theta\left(t\right){\tt Tr}\{\varrho_{\text{{tip}}}[f_{t}\left(t\right),f_{t}^{\dagger}\left(0\right)]_{+}\},\label{eq:PSI_R-1}
\end{align}
where $\varrho_{\text{{tip}}}$ is the density matrix of the system
described by the Hamiltonian in Eq. (\ref{eq:STM}).

Thus to determine an analytical expression for the LDOS, we apply
the equation-of-motion approach on Eq. (\ref{eq:PSI_R}). Such a procedure
is summarized as follows:

\begin{equation}
(\varepsilon+i\eta)\tilde{\mathcal{R}}_{\mathcal{AB}}=[\mathcal{A},\mathcal{B}^{\dagger}]_{+}+\tilde{\mathcal{R}}_{\left[\mathcal{A},\mathcal{\mathcal{H}}_{i}\right]\mathcal{B}},\label{eq:EOM}
\end{equation}
with $\eta\rightarrow0^{+}$, $\mathcal{A}$ and $\mathcal{B}$ as
fermionic operators belonging to the Hamiltonian $\mathcal{\mathcal{H}}_{i}$
($i=\text{host+ads}$ or tip).

By taking Eq. (\ref{eq:PSI_R}), one can calculate via Eqs. (\ref{eq:TIAM}),
(\ref{eq:PSI_R-1-1}) and (\ref{eq:EOM}) with $\mathcal{A}=\mathcal{B}=\psi_{0}$
and $\mathcal{\mathcal{H}}_{i}=\mathcal{H}_{\text{host+ads}}$, the
following relation
\begin{align}
\tilde{\mathcal{R}}_{\psi_{0}\psi_{0}} & =\mathcal{\tilde{\mathcal{R}}}_{f_{0}f_{0}}+(\pi\Gamma\rho_{0})q_{0}^{2}\sum_{jl}\tilde{\mathcal{R}}_{d_{j}d_{l}}+2(\pi\Gamma\rho_{0})^{1/2}q_{0}\nonumber \\
 & \times\sum_{j}\tilde{\mathcal{R}}_{d_{j}f_{0}},\label{eq:GF_1}
\end{align}
which depends on the Green's functions $\mathcal{\mathcal{\tilde{R}}}_{f_{0}f_{0}}$,
$\mathcal{\tilde{R}}_{f_{0}d_{j}}$ and $\mathcal{\tilde{R}}_{d_{j}d_{l}}$.
First, we find $\mathcal{\mathcal{\tilde{R}}}_{f_{0}f_{0}}$,

\begin{align}
\tilde{\mathcal{R}}_{f_{0}f{}_{0}} & =\pi\rho_{0}(\bar{\gamma}-i)+\pi\rho_{0}\Gamma(\bar{\gamma}-i)^{2}\sum_{jl}\tilde{\mathcal{R}}_{d_{j}d_{l}}\nonumber \\
\label{eq:g_ff-1}
\end{align}
and later on, the mixed Green's function $\tilde{\mathcal{R}}_{d_{j}f_{0}}$,
\begin{equation}
\tilde{\mathcal{R}}_{d_{j}f_{0}}=\sqrt{\pi\Gamma\rho_{0}}(\bar{\gamma}-i)\sum_{l}\tilde{\mathcal{R}}_{d_{j}d_{l}},\label{eq:g_df-1}
\end{equation}
where

\begin{equation}
\bar{\gamma}=\frac{1}{\pi\rho_{0}}\sum_{k}\frac{1}{\varepsilon-\varepsilon_{k}}.\label{eq:Fano_j}
\end{equation}

Now we choose for Eq. (\ref{eq:EOM}), $\mathcal{\mathcal{H}}_{i}=\mathcal{H}_{\text{tip}}$
and $\mathcal{A}=\mathcal{B}=f_{t}$, respectively, from Eqs. (\ref{eq:STM})
and (\ref{eq:STM_tip}), to show that

\begin{align}
\tilde{\mathcal{R}}_{f_{t}f_{t}} & =\pi\rho_{0}(\bar{\gamma}-i).\label{eq:tip}
\end{align}

In particular, for the wide band limit $D\rightarrow\infty$, $\bar{\gamma}\rightarrow0$.
Thus the imaginary parts of Eqs. (\ref{eq:g_ff-1}), (\ref{eq:g_df-1})
and (\ref{eq:tip}) become
\begin{align}
{\tt Im}(\tilde{\mathcal{R}}_{f_{0}f_{0}}) & =-\pi\rho_{0}[1+\Gamma\sum_{jl}{\tt Im}(\tilde{\mathcal{R}}_{d_{j}d_{l}})],\label{eq:I_G_ff-1}
\end{align}

\begin{align}
{\tt Im}(\tilde{\mathcal{R}}_{d_{j}f_{0}}) & =-\sqrt{\pi\Gamma\rho_{0}}\sum_{l}{\tt Re}(\tilde{\mathcal{R}}_{d_{j}d_{l}})\label{eq:ImMix}
\end{align}
and

\begin{equation}
{\tt Im}(\tilde{\mathcal{R}}_{f_{t}f_{t}})=-\pi\rho_{0}.\label{eq:Imtip}
\end{equation}

Now we take Eqs. (\ref{eq:I_G_ff-1}), (\ref{eq:ImMix}) and (\ref{eq:Imtip})
into Eq. (\ref{eq:transmit}) to obtain
\begin{align}
T_{\text{{probe}}}(\varepsilon) & =\frac{\mathcal{T}_{\text{{probe}}}(\varepsilon)}{\mathcal{T}_{b}}=1+\Gamma\sum_{jl}[(1-q_{0}^{2}){\tt Im}(\tilde{\mathcal{R}}_{d_{j}d_{l}})\nonumber \\
 & +2q_{0}{\tt Re}(\tilde{\mathcal{R}}_{d_{j}d_{l}})]\label{eq:LDOS_p1}
\end{align}
as the total transmittance through the system, expressed in terms
of the background conductance
\begin{equation}
\mathcal{T}_{b}=4x=4(\pi w\rho_{0})^{2}\label{eq:Tb1}
\end{equation}
and the Green's functions $\tilde{\mathcal{R}}_{d_{j}d_{l}}$ of the
adatoms.

\subsubsection{The STM tip in the same footing as the ``host+adatoms'' system}

\label{sub:2b3}

Here we derive the Landauer-Büttiker formula for the zero-bias conductance
$\mathcal{G}(0)$ by considering the STM tip in the same footing as
the ``host+adatoms'' system, which is achievable with $\tilde{V}=V$
in Eq. (\ref{eq:Fano_q}).

The zero-bias conductance is a function of the transmittance $\mathcal{T}_{\text{{full}}}\left(\varepsilon\right)$
as follows:

\begin{equation}
\mathcal{G}(0)=\frac{\partial}{\partial\varphi}\mathcal{J}_{\text{{host}}}(\varphi=0)=\frac{e^{2}}{h}\int d\varepsilon\left(-\frac{\partial f_{F}}{\partial\varepsilon}\right)\mathcal{T}_{\text{{full}}}(\varepsilon),\label{eq:_10b}
\end{equation}
with $\mathcal{J}_{\text{{host}}}$ as the current for the host and
$\mu_{\text{{host}}}-\mu_{\text{{tip}}}=e\varphi$, with $\varphi$
as the applied bias-voltage. We begin with the transformation

\begin{equation}
\left(\begin{array}{c}
c_{k}\\
b_{k}
\end{array}\right)=\left(\begin{array}{cc}
\frac{1}{\sqrt{2}} & \frac{1}{\sqrt{2}}\\
-\frac{1}{\sqrt{2}} & \frac{1}{\sqrt{2}}
\end{array}\right)\left(\begin{array}{c}
c_{ok}\\
c_{ek}
\end{array}\right)\label{eq:_01b}
\end{equation}
on the Hamiltonian of Eq. (\ref{eq:Total}), which depends on the
\textit{even} and \textit{odd} conduction operators $c_{ek}$ and
$c_{ok}$, respectively. These definitions allow us to express Eq.
(\ref{eq:Total}) as

\begin{equation}
\mathcal{H}=\mathcal{H}_{e}+\mathcal{H}_{o}+\mathcal{\tilde{H}}_{\text{{tun}}}=\mathcal{H}_{\varphi=0}+\mathcal{\tilde{H}}_{\text{{tun}}},\label{eq:H_total}
\end{equation}
where

\begin{eqnarray}
\mathcal{H}_{e} & = & \sum_{k}\varepsilon_{k}c_{ek}^{\dagger}c_{ek}+\sum_{j}\varepsilon_{j}d_{j}^{\dagger}d_{j}\nonumber \\
 & + & \sum_{jk}\sqrt{2}V(c_{ek}^{\dagger}d_{j}+\text{{H.c.}})+w\sum_{kq}c_{ek}^{\dagger}c_{eq}\nonumber \\
 & + & i\epsilon_{M}\eta_{1}\eta_{2}+\lambda(d_{1}-d_{1}^{\dagger})\eta_{1}\label{eq:even}
\end{eqnarray}
represents the Hamiltonian part of the system coupled to the adatoms
via an effective hybridization $\sqrt{2}V$, while

\begin{equation}
\mathcal{H}_{o}=\sum_{k}\varepsilon_{k}c_{ok}^{\dagger}c_{ok}-w\sum_{kq}c_{ok}^{\dagger}c_{oq}\label{eq:odd}
\end{equation}
is the decoupled one. However, they are connected to each other by
the tunneling Hamiltonian

\begin{equation}
\mathcal{\tilde{H}}_{\text{{tun}}}=-\Delta\mu\sum_{k}(c_{ek}^{\dagger}c_{ok}+c_{ok}^{\dagger}c_{ek}),\label{eq:tunn}
\end{equation}
with $\mu_{\text{{host}}}=\Delta\mu$, $\mu_{\text{{tip}}}=-\Delta\mu$
and $\Delta\mu=e\varphi/2$. As in the zero-bias regime $\Delta\mu\rightarrow0$,
due to $\varphi\rightarrow0$, $\mathcal{\tilde{H}}_{\text{{tun}}}$
is a perturbative term.

Here we use the interaction picture to calculate $\mathcal{T}_{\text{{full}}}(\varepsilon)$.
It ensures that a state $\left|\Phi_{n}\right\rangle $ from the spectrum
of the Hamiltonian given by Eq. (\ref{eq:H_total}) admits the following
time-dependency

\begin{align}
\left|\Phi_{n}\right\rangle  & =e^{-\frac{i}{\hbar}\int_{-\infty}^{0}\mathcal{\tilde{H}}_{\text{{tun}}}(\tau)d\tau}\left|\Psi_{n}\right\rangle \nonumber \\
 & \simeq(1-\frac{i}{\hbar}\int_{-\infty}^{0}\mathcal{\tilde{H}}_{\text{{tun}}}(\tau)d\tau)\left|\Psi_{n}\right\rangle ,\label{eq:phi}
\end{align}
where $\hbar=\frac{h}{2\pi}$ and $\left|\Psi_{n}\right\rangle $
is an eigenstate of $\mathcal{H}_{e}+\mathcal{H}_{o}=\mathcal{H}_{\varphi=0}$.
Thus the current $\mathcal{J}_{\text{{host}}}$ for the host can be
obtained by performing the expected mean value of the current operator
$\mathcal{I}_{\text{{host}}}\equiv\mathcal{I}_{\text{{host}}}\left(t=0\right)$,
which reads

\begin{align}
\mathcal{J}_{\text{{host}}} & =\left\langle \Phi_{n}\right|\mathcal{I}_{\text{{host}}}\left|\Phi_{n}\right\rangle \nonumber \\
 & =-\frac{i}{\hbar}\left\langle \Psi_{n}\right|\int_{-\infty}^{0}[\mathcal{I}_{\text{{host}}},\mathcal{\tilde{H}}_{\text{{tun}}}(\tau)]d\tau\left|\Psi_{n}\right\rangle +\mathcal{O}(\mathcal{\tilde{H}}_{\text{{tun}}}^{2}),\label{eq:current}
\end{align}
where we have regarded $\left\langle \Psi_{n}\right|\mathcal{I}_{\text{{host}}}\left|\Psi_{n}\right\rangle =0$
and by considering the thermal average on the latter equation, which
gives

\begin{eqnarray}
\mathcal{J}_{\text{{host}}} & = & -\frac{i}{\hbar}\int_{-\infty}^{0}{\tt Tr}\{\varrho_{\varphi=0}[\mathcal{I}_{\text{{host}}},\mathcal{\tilde{H}}_{\text{{tun}}}(\tau)]\}d\tau,\label{eq:_07c}
\end{eqnarray}
where $\varrho_{\varphi=0}$ is the density matrix of the system described
by the Hamiltonian $\mathcal{H}_{\varphi=0}$ in Eq. (\ref{eq:H_total}).
By applying the equation-of-motion on $\mathcal{I}_{\text{{host}}}$,
we show that

\begin{eqnarray}
\mathcal{I}_{\text{{host}}} & = & -\frac{i}{\hbar}[e\sum_{k}c_{k}^{\dagger}c_{k},\mathcal{H}_{\varphi=0}]\nonumber \\
 & = & \left(-\frac{ie}{\sqrt{2}\hbar}\right)V\sum_{kj}\left\{ (c_{ek}^{\dagger}d_{j}-d_{j}^{\dagger}c_{ek})\right.\nonumber \\
 & + & \left.(c_{ok}^{\dagger}d_{j}-d_{j}^{\dagger}c_{ok})\right\} \nonumber \\
 & + & \left(-\frac{ie}{\hbar}\right)w\sum_{q\tilde{q}}(c_{oq}^{\dagger}c_{e\tilde{q}}-c_{e\tilde{q}}^{\dagger}c_{oq}),\label{eq:11b}
\end{eqnarray}
which, in combination with Eq. (\ref{eq:_07c}), leads to
\begin{align}
\mathcal{J}_{\text{{host}}} & =-\frac{e}{\hbar}\Delta\mu{\tt Im}\int_{-\infty}^{+\infty}d\tau\{\sqrt{2}V\sum_{j}\mathcal{F}_{j}(-\tau)\nonumber \\
 & +2w\mathcal{M}(-\tau)\},\label{eq:J_B}
\end{align}
where

\begin{align}
\mathcal{F}_{j}(-\tau) & =-\frac{i}{\hbar}\theta(-\tau){\tt Tr}\{\varrho_{\varphi=0}[f_{o}^{\dagger}d_{j},\sum_{q}c_{eq}^{\dagger}(\tau)c_{oq}(\tau)]\}\label{eq:F}
\end{align}
and

\begin{align}
\mathcal{M}(-\tau) & =-\frac{i}{\hbar}\theta(-\tau){\tt Tr}\{\varrho_{\varphi=0}[f_{o}^{\dagger}f_{e},\sum_{k}c_{ek}^{\dagger}(\tau)c_{ok}(\tau)]\}\label{eq:M}
\end{align}
are retarded Green's functions, expressed in terms of the operators

\begin{equation}
f_{o}=\sum_{\tilde{q}}c_{o\tilde{q}}\label{eq:25a}
\end{equation}
and
\begin{equation}
f_{e}=\sum_{q}c_{eq}.\label{eq:25b}
\end{equation}

In order to find a closed expression for the current $\mathcal{J}_{\text{{host}}}$,
we should evaluate the integrals in the time coordinate $\tau$ of
Eq. (\ref{eq:J_B}), which result in

\begin{eqnarray}
\int_{-\infty}^{+\infty}d\tau\mathcal{F}_{j}(-\tau) & = & \mathbb{\mathcal{Z}}^{-1}\sum_{mn}\frac{(e^{-\beta E_{n}}-e^{-\beta E_{m}})}{E_{n}-E_{m}+i\eta}\nonumber \\
 & \times & \left\langle \Psi_{n}\right|f_{o}^{\dagger}d_{j}\left|\Psi_{m}\right\rangle \left\langle \Psi_{m}\right|\sum_{q}c_{eq}^{\dagger}c_{oq}\left|\Psi_{n}\right\rangle \nonumber \\
\label{eq:int_1}
\end{eqnarray}
and

\begin{eqnarray}
\int_{-\infty}^{+\infty}d\tau\mathcal{\mathcal{M}}(-\tau) & = & \mathbb{\mathcal{Z}}^{-1}\sum_{mn}\frac{(e^{-\beta E_{n}}-e^{-\beta E_{m}})}{E_{n}-E_{m}+i\eta}\nonumber \\
 & \times & \left\langle \Psi_{n}\right|f_{o}^{\dagger}f_{e}\left|\Psi_{m}\right\rangle \left\langle \Psi_{m}\right|\sum_{q}c_{eq}^{\dagger}c_{oq}\left|\Psi_{n}\right\rangle ,\nonumber \\
\label{eq:int_2}
\end{eqnarray}
where we have used $\mathbb{\mathcal{Z}}$ as the partition function
of $\mathcal{H}_{\varphi=0}\left|\Psi_{m}\right\rangle =E_{m}\left|\Psi_{m}\right\rangle $,
$\mathcal{A}\left(\tau\right)=e^{\frac{i}{\hbar}\mathcal{H}_{\varphi=0}\tau}\mathcal{A}e^{-\frac{i}{\hbar}\mathcal{H}_{\varphi=0}\tau}$
for an arbitrary time-dependent operator $\mathcal{A}\left(\tau\right)$
and $\eta\rightarrow0^{+}$. To eliminate the matrix element $\left\langle \Psi_{m}\right|c_{eq}^{\dagger}c_{oq}\left|\Psi_{n}\right\rangle $
in Eqs. (\ref{eq:int_1}) and (\ref{eq:int_2}), we calculate $\left\langle \Psi_{m}\right|[\sum_{q}c_{eq}^{\dagger}c_{oq},\mathcal{H}_{\varphi=0}]\left|\Psi_{n}\right\rangle $,
which gives

\begin{eqnarray}
\left\langle \Psi_{m}\right|\sum_{q}c_{eq}^{\dagger}c_{oq}\left|\Psi_{n}\right\rangle  & = & -\frac{\sqrt{2}V}{(E_{n}-E_{m})}\nonumber \\
 & \times & \sum_{\tilde{j}}\left\langle \Psi_{m}\right|d_{\tilde{j}}^{\dagger}f_{o}\left|\Psi_{n}\right\rangle \nonumber \\
 & - & \frac{2w}{(E_{n}-E_{m})}\left\langle \Psi_{m}\right|f_{e}^{\dagger}f_{o}\left|\Psi_{n}\right\rangle .\nonumber \\
\label{eq:mix}
\end{eqnarray}

By performing the substitutions of Eqs. (\ref{eq:int_1}), (\ref{eq:int_2})
with (\ref{eq:mix}) in Eq. (\ref{eq:J_B}), we enclose the result
into the function labeled by $\chi_{mn}$ to show that
\begin{align}
\mathcal{J}_{\text{{host}}} & =\frac{e}{\hbar}\pi\Delta\mu\mathbb{\mathcal{Z}}^{-1}\sum_{mn}\chi_{mn}\frac{(e^{-\beta E_{n}}-e^{-\beta E_{m}})}{E_{n}-E_{m}}\delta(E_{n}-E_{m})\nonumber \\
 & =-\frac{e}{\hbar}\pi\Delta\mu\beta\sum_{mn}[\mathbb{\mathcal{Z}}^{-1}e^{-\beta E_{n}}\delta(E_{n}-E_{m})]\chi_{nm},\label{eq:J_B_2}
\end{align}
where we have defined
\begin{multline}
\chi_{nm}=(\sqrt{2}V)^{2}\sum_{j\tilde{j}}\left\langle \Psi_{n}\right|f_{o}^{\dagger}d_{j}\left|\Psi_{m}\right\rangle \left\langle \Psi_{m}\right|d_{\tilde{j}}^{\dagger}f_{o}\left|\Psi_{n}\right\rangle \\
+2\sqrt{2}V(2w)\sum_{j}\left\langle \Psi_{n}\right|f_{o}^{\dagger}d_{j}\left|\Psi_{m}\right\rangle \left\langle \Psi_{m}\right|f_{e}^{\dagger}f_{o}\left|\Psi_{n}\right\rangle \\
+(2w)^{2}\left\langle \Psi_{n}\right|f_{o}^{\dagger}f_{e}\left|\Psi_{m}\right\rangle \left\langle \Psi_{m}\right|f_{e}^{\dagger}f_{o}\left|\Psi_{n}\right\rangle .\label{eq:qui_mn}
\end{multline}

In this calculation we have used
\begin{multline*}
\left\langle \Psi_{n}\right|f_{o}^{\dagger}d_{j}\left|\Psi_{m}\right\rangle \left\langle \Psi_{m}\right|f_{e}^{\dagger}f_{o}\left|\Psi_{n}\right\rangle \\
=\left\langle \Psi_{n}\right|f_{o}^{\dagger}f_{e}\left|\Psi_{m}\right\rangle \left\langle \Psi_{m}\right|d_{j}^{\dagger}f_{o}\left|\Psi_{n}\right\rangle ,
\end{multline*}
with
\begin{equation}
\frac{(e^{-\beta E_{n}}-e^{-\beta E_{m}})}{E_{n}-E_{m}}=-\beta e^{-\beta E_{n}}\label{eq:_limit}
\end{equation}
in the limit $E_{n}\rightarrow E_{m}$. The property $\left[\mathcal{H}_{e},\mathcal{H}_{o}\right]=0$
ensures the partitions $E_{n}=E_{n}^{e}+E_{n}^{o}$ and $\mathbb{\mathcal{Z}}=\mathbb{\mathcal{Z}}_{e}\mathbb{\mathcal{Z}}_{o}$
for the Hamiltonians $\mathcal{H}_{e}$ and $\mathcal{H}_{o}$, respectively
in the brackets of Eq. (\ref{eq:J_B_2}), thus leading to

\begin{align}
 & \mathbb{\mathcal{Z}}^{-1}e^{-\beta E_{n}}\delta(E_{n}-E_{m})=\frac{1}{\beta}\mathbb{\mathcal{Z}}_{e}^{-1}\mathbb{\mathcal{Z}}_{o}^{-1}\int d\varepsilon\left(-\frac{\partial f_{F}}{\partial\varepsilon}\right)\nonumber \\
 & \times(e^{-\beta E_{n}^{e}}+e^{-\beta E_{m}^{e}})(e^{-\beta E_{n}^{o}}+e^{-\beta E_{m}^{o}})\delta(\varepsilon+E_{n}^{e}-E_{m}^{e})\nonumber \\
 & \times\delta(\varepsilon+E_{n}^{o}-E_{m}^{o}).\label{eq:prop}
\end{align}

Therefore, we substitute Eqs. (\ref{eq:qui_mn}) and (\ref{eq:prop})
in Eq. (\ref{eq:J_B_2}) to calculate $\frac{\partial}{\partial\varphi}\mathcal{J}_{\text{{host}}}(\varphi=0)$.
The comparison of such a result with Eq. (\ref{eq:_10b}) allows us
to find

\begin{equation}
\mathcal{T}_{\text{{full}}}\left(\varepsilon\right)=(2\pi w)^{2}\tilde{\rho}_{\text{LDOS}}(\varepsilon)\tilde{\rho}_{\text{tip}}(\varepsilon),\label{eq:transmit-1}
\end{equation}
where

\begin{equation}
\tilde{\rho}_{\text{LDOS}}(\varepsilon)=-\frac{1}{\pi}{\tt Im}(\tilde{\mathcal{R}}_{\psi_{e}\psi_{e}})\label{eq:FM_LDOS-1}
\end{equation}
and

\begin{align}
\mathcal{R}_{\psi_{e}\psi_{e}} & =-\frac{i}{\hbar}\theta\left(t\right){\tt Tr}\{\varrho_{\text{e}}[\psi_{e}\left(t\right),\psi_{e}^{\dagger}\left(0\right)]_{+}\},\label{eq:PSI_R-2}
\end{align}
with the former as the renormalized LDOS of the ``host+adatoms''
system described by the Hamiltonian of Eq. (\ref{eq:even}), which
is affected by the STM tip via the scattering term $w\sum_{kq}c_{ek}^{\dagger}c_{eq}$,
thus leading to

\begin{equation}
\psi_{e}=f_{e}+(\pi\Delta\rho_{0})^{1/2}\gamma\sum_{j}d_{j}\label{eq:Psi_e}
\end{equation}
and

\begin{align*}
\tilde{\mathcal{R}}_{\psi_{e}\psi_{e}} & =\tilde{\mathcal{R}}_{f_{e}f_{e}}+(\pi\rho_{0}\Delta)\gamma^{2}\sum_{jl}\tilde{\mathcal{R}}_{d_{j}d_{l}}+2(\pi\rho_{0}\Delta)^{1/2}\gamma\\
 & \times\sum_{j}\tilde{\mathcal{R}}_{d_{j}f_{e}}
\end{align*}
that generalize Eqs. (\ref{eq:PSI_R-1-1}) and (\ref{eq:GF_1}),
respectively, with a renormalized Anderson parameter

\begin{equation}
\Delta=2\pi V^{2}\rho_{0}\label{eq:AP_2}
\end{equation}
and Fano factor

\begin{equation}
\gamma=\left(\pi\rho_{0}\Delta\right)^{-1/2}\left(\frac{\sqrt{2}V}{2w}\right).\label{eq:Fano-1}
\end{equation}

Additionally, the scattering term $-w\sum_{kq}c_{ok}^{\dagger}c_{oq}$
renormalizes the DOS of the STM tip due to the Hamiltonian of Eq.
(\ref{eq:odd}), which provides

\begin{equation}
\tilde{\rho}_{\text{tip}}(\varepsilon)=-\frac{1}{\pi}{\tt Im}(\tilde{\mathcal{R}}_{f_{o}f_{o}}).\label{eq:STM_tip-1}
\end{equation}

From Eqs. (\ref{eq:even}) and (\ref{eq:25b}), we make the substitutions
$\mathcal{A}=\mathcal{B}=f{}_{e}$ and $\mathcal{H}_{i}=\mathcal{H}_{e}$
in Eq. (\ref{eq:EOM}), which gives

\begin{align}
\tilde{\mathcal{R}}_{f_{e}f{}_{e}} & =\frac{\pi\rho_{0}(\bar{\gamma}-i)}{1-\sqrt{x}(\bar{\gamma}-i)}+\pi\rho_{0}\Delta\left[\frac{(\bar{\gamma}-i)}{1-\sqrt{x}(\bar{\gamma}-i)}\right]^{2}\nonumber \\
 & \times\sum_{jl}\tilde{\mathcal{R}}_{d_{j}d_{l}}\left(\varepsilon\right),\label{eq:g_ff}
\end{align}
where we have used the mixed Green's function
\begin{equation}
\tilde{\mathcal{R}}_{d_{j}f_{e}}=\sqrt{\pi\Delta\rho_{0}}\frac{(\bar{\gamma}-i)}{1-\sqrt{x}(\bar{\gamma}-i)}\sum_{l}\tilde{\mathcal{R}}_{d_{j}d_{l}},\label{eq:g_df}
\end{equation}
determined from Eq. (\ref{eq:EOM}) by considering $\mathcal{A}=d_{j}$,
$\mathcal{B}=f{}_{e}$ and $\mathcal{H}_{i}=\mathcal{H}_{e}$, with
the parameter $x$ being the same as found in Eq. (\ref{eq:Tb1}).
We point out that, Eqs. (\ref{eq:g_ff}) and (\ref{eq:g_df}), constitute
respectively, generalizations of Eqs. (\ref{eq:g_ff-1}) and (\ref{eq:g_df-1}),
where the latter can be obtained from the former by making $x\ll1$.
Thus, the imaginary parts of Eqs. (\ref{eq:g_ff}) and (\ref{eq:g_df})
for the wide band limit $D\rightarrow\infty$, become

\begin{align}
{\tt Im}(\tilde{\mathcal{R}}_{f_{e}f_{e}}) & =-\frac{\pi\rho_{0}}{1+x}-\frac{(1-x)}{(1+x)^{2}}\pi\Delta\rho_{0}\sum_{jl}{\tt Im}(\tilde{\mathcal{R}}_{d_{j}d_{l}})\nonumber \\
 & +\frac{2\sqrt{x}}{\left(1+x\right)^{2}}\pi\Delta\rho_{0}\sum_{jl}{\tt Re}(\tilde{\mathcal{R}}_{d_{j}d_{l}})\label{eq:I_G_ff}
\end{align}
and
\begin{align}
{\tt Im}(\tilde{\mathcal{R}}_{d_{j}f_{e}}) & =-\frac{\sqrt{x\pi\Delta\rho_{0}}}{1+x}\sum_{l}{\tt Im}(\tilde{\mathcal{R}}_{d_{j}d_{l}})\nonumber \\
 & -\frac{\sqrt{\pi\Delta\rho_{0}}}{1+x}\sum_{l}{\tt Re}(\tilde{\mathcal{R}}_{d_{j}d_{l}}),\label{eq:I_G_df}
\end{align}
where we have used $\bar{\gamma}\rightarrow0$. To conclude, we notice
that $\tilde{\mathcal{R}}_{f_{o}f_{o}}$ is decoupled from the adatoms.
Thereby, from Eqs. (\ref{eq:odd}) and (\ref{eq:25a}), we take $\mathcal{A}=\mathcal{B}=f{}_{o}$
and $\mathcal{H}_{i}=\mathcal{H}_{o}$ in Eq. (\ref{eq:EOM}) and
we obtain
\begin{equation}
{\tt Im}(\tilde{\mathcal{R}}_{f_{o}f_{o}})=-\frac{\pi\rho_{0}}{1+x},\label{eq:rho_ff}
\end{equation}
which is equal to the first term of Eq. (\ref{eq:I_G_ff}).

Thus the substitution of Eqs. (\ref{eq:I_G_ff}), (\ref{eq:I_G_df}),
and (\ref{eq:rho_ff}) in Eq. (\ref{eq:transmit-1}), leads to
\begin{align}
T_{\text{{full}}}(\varepsilon) & =\frac{\mathcal{T}_{\text{{full}}}(\varepsilon)}{\bar{\mathcal{T}}_{b}}=1+\bar{\Gamma}\sum_{jl}[(1-q_{b}^{2}){\tt Im}(\tilde{\mathcal{R}}_{d_{j}d_{l}})\nonumber \\
 & +2q_{b}{\tt Re}(\tilde{\mathcal{R}}_{d_{j}d_{l}})],\label{eq:trans2}
\end{align}
where

\begin{equation}
\bar{\mathcal{T}}_{b}=\frac{4x}{\left(1+x\right)^{2}}\label{eq:d2-1-1}
\end{equation}
represents the transmittance in the absence of the adatoms and MBSs
(background contribution),

\begin{equation}
\bar{\Gamma}=\frac{\Delta}{1+x}\label{eq:d4-1-1}
\end{equation}
is an effective coupling and

\begin{equation}
q_{b}=\frac{\left(1-x\right)}{2\sqrt{x}}\label{eq:Fano_qb}
\end{equation}
is the Fano parameter. Notice that Eq. (\ref{eq:trans2}) has the
same form of Eq. (\ref{eq:LDOS_p1}), but with $q_{0}$ replaced by
$q_{b}$. In this work, we will focus on the limit $q_{0}=q_{b}=0$.

\subsection{Green's functions of the adatoms}

In this section, we calculate the Green's functions $\tilde{\mathcal{R}}_{d_{j}d_{l}}$
within the wide band limit $D\rightarrow\infty$. We point out that,
the expressions derived here describe the situation of Sec. \ref{sub:2b2}
for an STM tip as probe with $\Gamma$ instead of $\Delta$ {[}see
Eqs. (\ref{eq:AP}) and (\ref{eq:AP_2}){]} and by assuming $x\ll1$
{[}Eq. (\ref{eq:Tb1}){]}, otherwise, they belong to the case of Sec.
\ref{sub:2b3}. We begin by applying the equation-of-motion procedure
on

\begin{align}
\mathcal{R}_{d_{j}d_{l}} & =-\frac{i}{\hbar}\theta\left(t\right){\tt Tr}\{\varrho_{\text{{s}}}[d_{j}\left(t\right),d_{l}^{\dagger}\left(0\right)]_{+}\},\label{eq:djdl}
\end{align}
where $\text{{s=host+ads}}$ or $\text{{s=e}, }$and changing to the
energy domain $\varepsilon$, we obtain the following relation:

\begin{align}
(\varepsilon-\tilde{\varepsilon}_{j}-i\Sigma^{I}-\delta_{j1}\Sigma_{\text{{MBS1}}})\tilde{\mathcal{R}}_{d_{j}d_{l}} & =\delta_{jl}+\Sigma\sum_{\tilde{l}\neq j}\tilde{\mathcal{R}}_{d_{\tilde{l}}d_{l}},\label{eq:GjjA}
\end{align}
with

\begin{equation}
\tilde{\varepsilon}_{j}=\varepsilon_{j}+\Sigma^{R}\label{eq:rnej}
\end{equation}
is the adatom level renormalized by the STM tip-host coupling $w$,
with $\Sigma=\Sigma^{R}+i\Sigma^{I}$,

\begin{equation}
\Sigma^{R}=-\frac{\sqrt{x}}{1+x}\Delta,\label{eq:setil}
\end{equation}

\begin{equation}
\Sigma^{I}=-\frac{\Delta}{1+x}\label{eq:gtil}
\end{equation}
and

\begin{equation}
\Sigma_{\text{{MBS1}}}=\lambda^{2}K(1+\lambda^{2}\tilde{K})\label{eq:self_MBS}
\end{equation}
as the self-energy due to the MBS 1 coupled to the adatom 1,

\begin{equation}
K=\frac{1}{2}\left(\frac{1}{\varepsilon-\epsilon_{M}+i\eta}+\frac{1}{\varepsilon+\epsilon_{M}+i\eta}\right)\label{eq:KE_1}
\end{equation}
and

\begin{equation}
\tilde{K}=\frac{K}{\varepsilon+\tilde{\varepsilon}_{1}-i\Sigma^{I}-\lambda^{2}K},\label{eq:cm17-1}
\end{equation}
which have the same forms as found in Ref. {[}\onlinecite{key-25}{]}.
Thus the solution of Eq. (\ref{eq:GjjA}) provides

\begin{align}
\tilde{\mathcal{R}}_{d_{1}d_{1}} & =\frac{1}{\varepsilon-\tilde{\varepsilon}_{1}-i\Sigma^{I}-\Sigma_{\text{{MBS1}}}-\mathcal{C}_{2}}\nonumber \\
\label{eq:d1d1}
\end{align}
as the Green's function of the adatom 1, with

\begin{equation}
\mathcal{C}_{j}=\frac{(\Sigma^{R}+i\Sigma^{I})^{2}}{\varepsilon-\tilde{\varepsilon}_{j}-i\Sigma^{I}},\label{eq:G}
\end{equation}
as the self-energy due to the presence of the \textit{jth} adatom.
For $\mathcal{C}_{2}=0$, we highlight that Eq. (\ref{eq:d1d1}) is
reduced to the Green's function of the single QD system found in Ref.
{[}\onlinecite{key-25}{]}. In the case of the adatom 2, we have

\begin{align}
\tilde{\mathcal{R}}_{d_{2}d_{2}} & =\frac{1-\tilde{\mathcal{R}}_{d_{1}d_{1}}^{0}\Sigma_{\text{{MBS1}}}}{\varepsilon-\tilde{\varepsilon}_{2}-i\Sigma^{I}-\dfrac{\tilde{\mathcal{R}}_{d_{1}d_{1}}^{0}}{\tilde{\mathcal{R}}_{d_{2}d_{2}}^{0}}\Sigma_{\text{{MBS1}}}-\mathcal{C}_{1}},\label{eq:d2d2}
\end{align}
where $\tilde{\mathcal{R}}_{d_{1}d_{1}}^{0}=1/(\varepsilon-\tilde{\varepsilon}_{1}-i\Sigma^{I})$
and $\tilde{\mathcal{R}}_{d_{2}d_{2}}^{0}=1/(\varepsilon-\tilde{\varepsilon}_{2}-i\Sigma^{I})$
represent the corresponding Green's functions for the single adatom
system without Majoranas. To conclude, the mixed Green's functions
are

\begin{equation}
\tilde{\mathcal{R}}_{d_{2}d_{1}}=\frac{\Sigma^{R}+i\Sigma^{I}}{\varepsilon-\tilde{\varepsilon}_{2}-i\Sigma^{I}}\tilde{\mathcal{R}}_{d_{1}d_{1}}\label{eq:d2d1}
\end{equation}
and

\begin{align}
\tilde{\mathcal{R}}_{d_{1}d_{2}} & =\frac{\Sigma^{R}+i\Sigma^{I}}{\varepsilon-\tilde{\varepsilon}_{1}-i\Sigma^{I}-\Sigma_{\text{{MBS1}}}}\tilde{\mathcal{R}}_{d_{2}d_{2}}.\label{eq:d1d2}
\end{align}

The main result of this section is the emergence of a lack of symmetry
in these Green's functions. This property lies on the coupling of
the adatom 1 with the MBS 1. To notice that, let us examine the situation
where the adatom 1 is decoupled from the MBS 1, which can be obtained
with $\Sigma_{\text{{MBS1}}}=0$ in Eq. (\ref{eq:self_MBS}). By inspection
of Eqs. (\ref{eq:d1d1}), (\ref{eq:d2d2}), (\ref{eq:d2d1}) and (\ref{eq:d1d2}),
we verify that the functions $\tilde{\mathcal{R}}_{d_{1}d_{1}}$ and
$\tilde{\mathcal{R}}_{d_{1}d_{2}}$ can be determined by the swap
of the indexes $1\leftrightarrow2$ in $\tilde{\mathcal{R}}_{d_{2}d_{2}}$
and $\tilde{\mathcal{R}}_{d_{2}d_{1}}$, respectively. However, in
the opposite situation with $\Sigma_{\text{{MBS1}}}\neq0$, this symmetry
is broken. Thus in Sec. \ref{sec3}, we will investigate this lack
of symmetry via the transmittances of Eqs. (\ref{eq:LDOS_p1}) and
(\ref{eq:trans2}). To this end, we will follow the two-stage procedure
presented in Sec. \ref{sec:intro}.

\section{Results}

\label{sec3}

Here we consider the Kitaev chain long enough, which forces $\epsilon_{M}\sim e^{-L/\xi}\rightarrow0$
in Eq. (\ref{eq:TIAM}) for $L\gg\xi$. We adopt typical values for
adatoms in metals {[}Ref. \onlinecite{{AHCNeto}}{]}: $\Delta=\Gamma=0.2$
for the Anderson parameters of Eqs. (\ref{eq:AP}) and (\ref{eq:AP_2}),
$\lambda$, $\varepsilon_{1}=-\frac{\Delta\varepsilon}{2}$, $\varepsilon_{2}=\frac{\Delta\varepsilon}{2}$,
the symmetric detuning $\Delta\varepsilon=\varepsilon_{2}-\varepsilon_{1}$
and $\varepsilon$ in units of eV.

\begin{figure}[h]
\centerline{\resizebox{3.5in}{!}{\includegraphics[clip,width=0.14\textwidth,height=0.07\textheight]{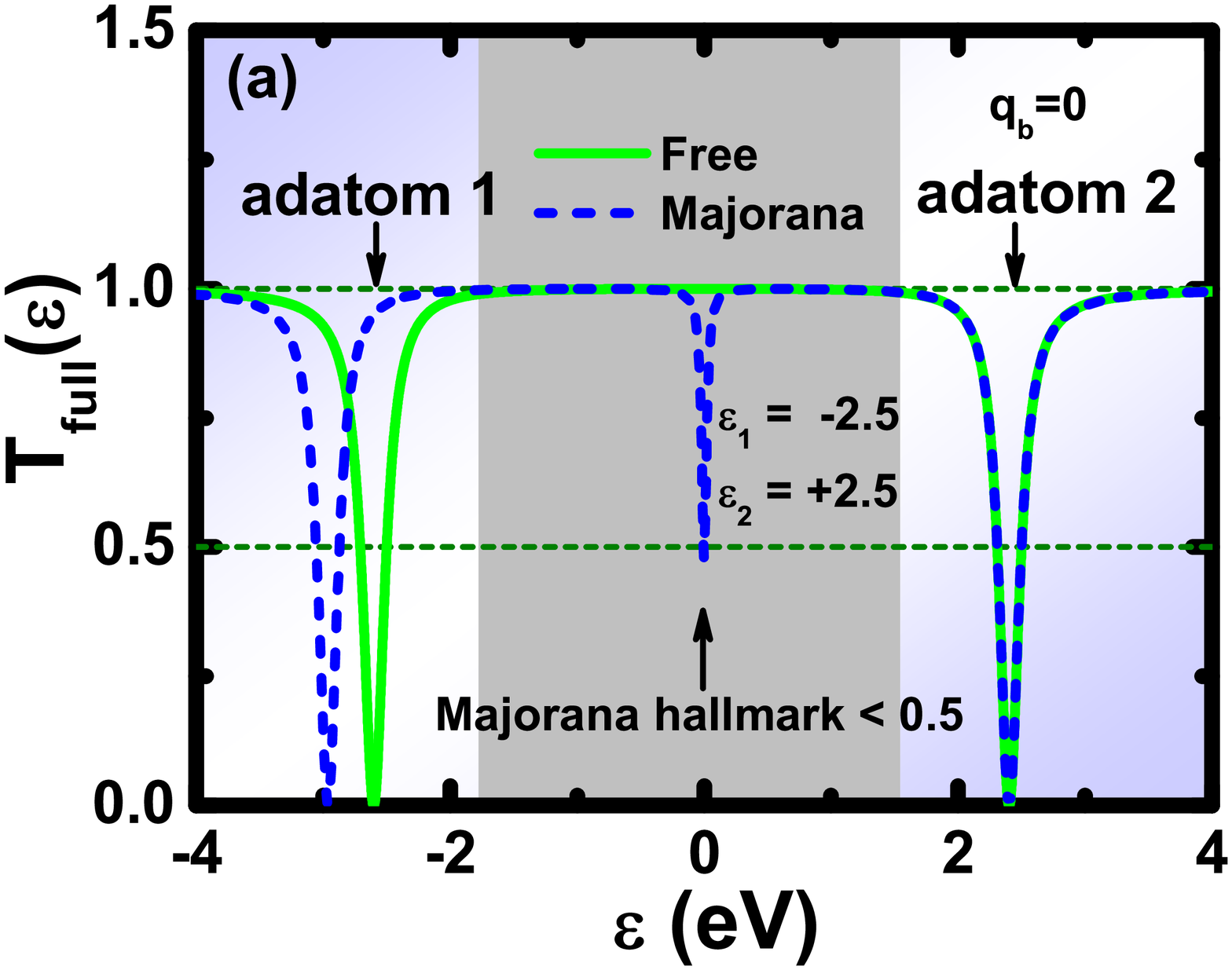}}}\centerline{\resizebox{3.5in}{!}{\includegraphics[clip,width=0.14\textwidth,height=0.07\textheight]{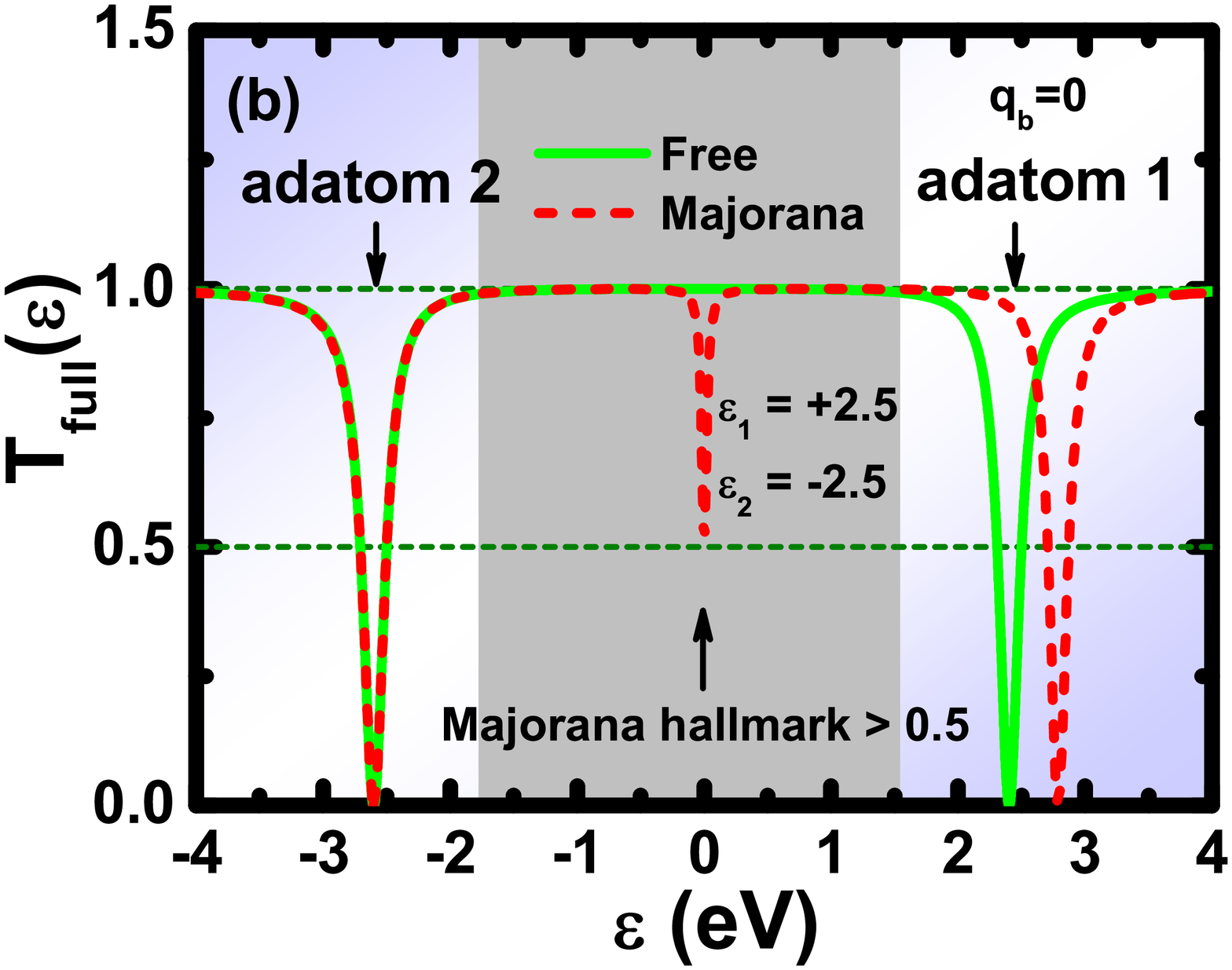}}}
\caption{(Color online) Parameters employed: $\epsilon_{M}=0$ {[}long enough
Kitaev chain{]}, $\lambda=5\Delta$ and $\Delta=0.2$ {[}see Eqs.
(\ref{eq:TIAM}) and (\ref{eq:AP_2}){]}. Transmittance $T_{\text{{full}}}(\varepsilon)$
determined by Eq. (\ref{eq:trans2}) in the Fano regime $q_{b}=0$
{[}Eq. (\ref{eq:Fano_qb}){]} as a function of the single particle
energy $\varepsilon$. In the panels (a) and (b) we have: the solid-green
lineshape is for the apparatus of Fig. (\ref{Fig1}) in the absence
of the MBS 1. Implementation of the two-stage procedure of Sec. \ref{sec:intro}:
(a) $\varepsilon_{1}=-2.5$ and $\varepsilon_{2}=+2.5$: The dashed-blue
curve corresponds to the system coupled to the MBS 1. (b) $\varepsilon_{1}=+2.5$
and $\varepsilon_{2}=-2.5$: The dashed-red lineshape is for the MBS
1. Here we see the main result of this procedure: the formation of
a Majorana dip with an amplitude that fluctuates slightly around $1/2$,
but it remains pinned at zero-bias even by performing the gates swap.
The satellite dips do not share such a feature, they become significantly
shifted under the permutation of the levels in the adatoms. }

\label{Fig2}
\end{figure}

In order to investigate the transmittance $T_{\text{{full}}}(\varepsilon)$
of Eq. (\ref{eq:trans2}) as a function of the single particle energy
$\varepsilon$, in Fig. \ref{Fig2} we use $\lambda=5\Delta$, with
$x=1$ and Fano factor $q_{b}=0$ {[}Eq. (\ref{eq:Fano_qb}){]}. This
set of parameters allows one to emulate the situation where the STM
tip is strongly connected to the host surface and therefore, considered
in the same footing as the ``host+adatoms'' system.

\begin{figure}[h]
\centerline{\resizebox{3.5in}{!}{\includegraphics[clip,width=0.145\textwidth,height=0.07\textheight]{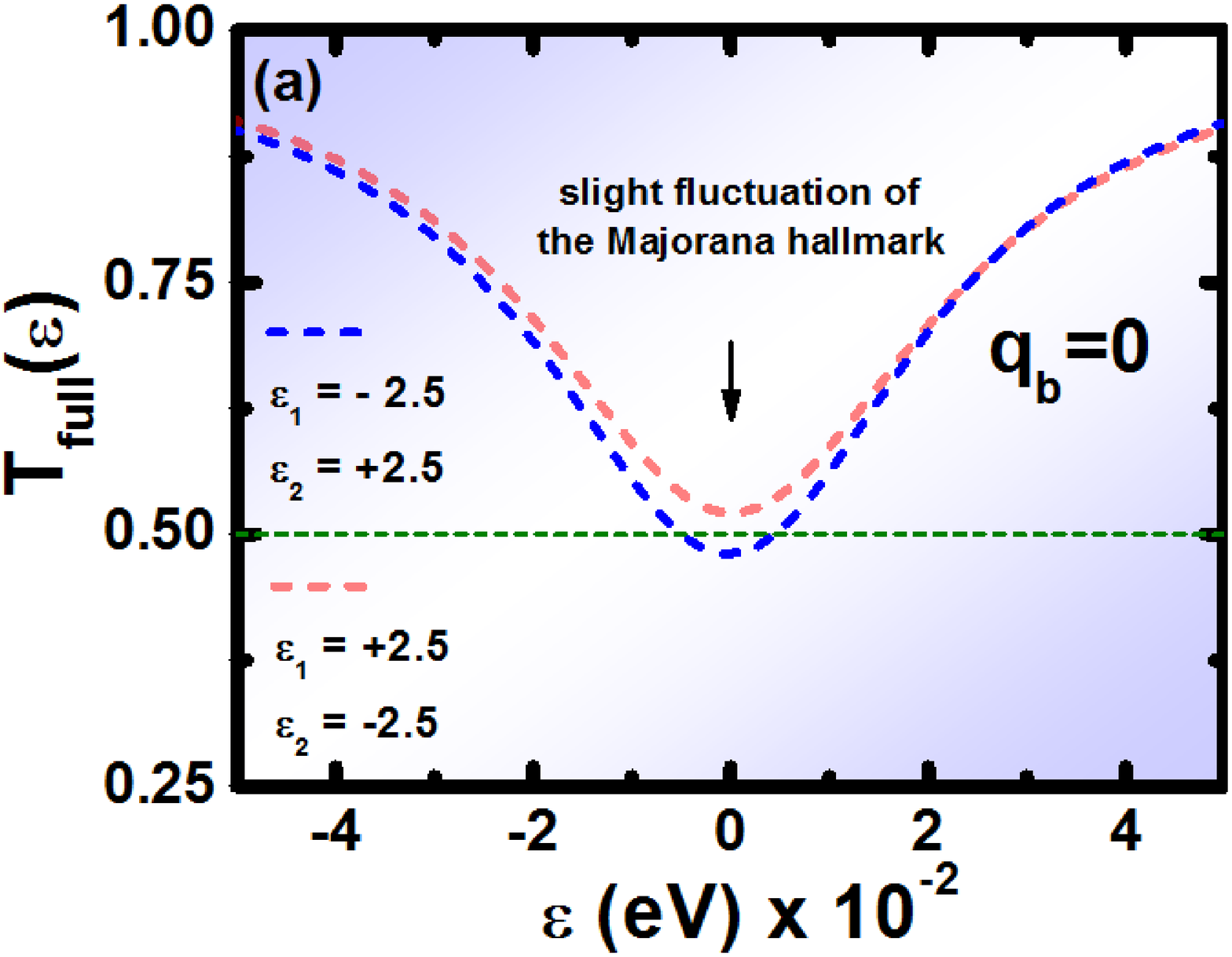}}}\centerline{\resizebox{3.5in}{!}{\includegraphics[clip,width=0.14\textwidth,height=0.07\textheight]{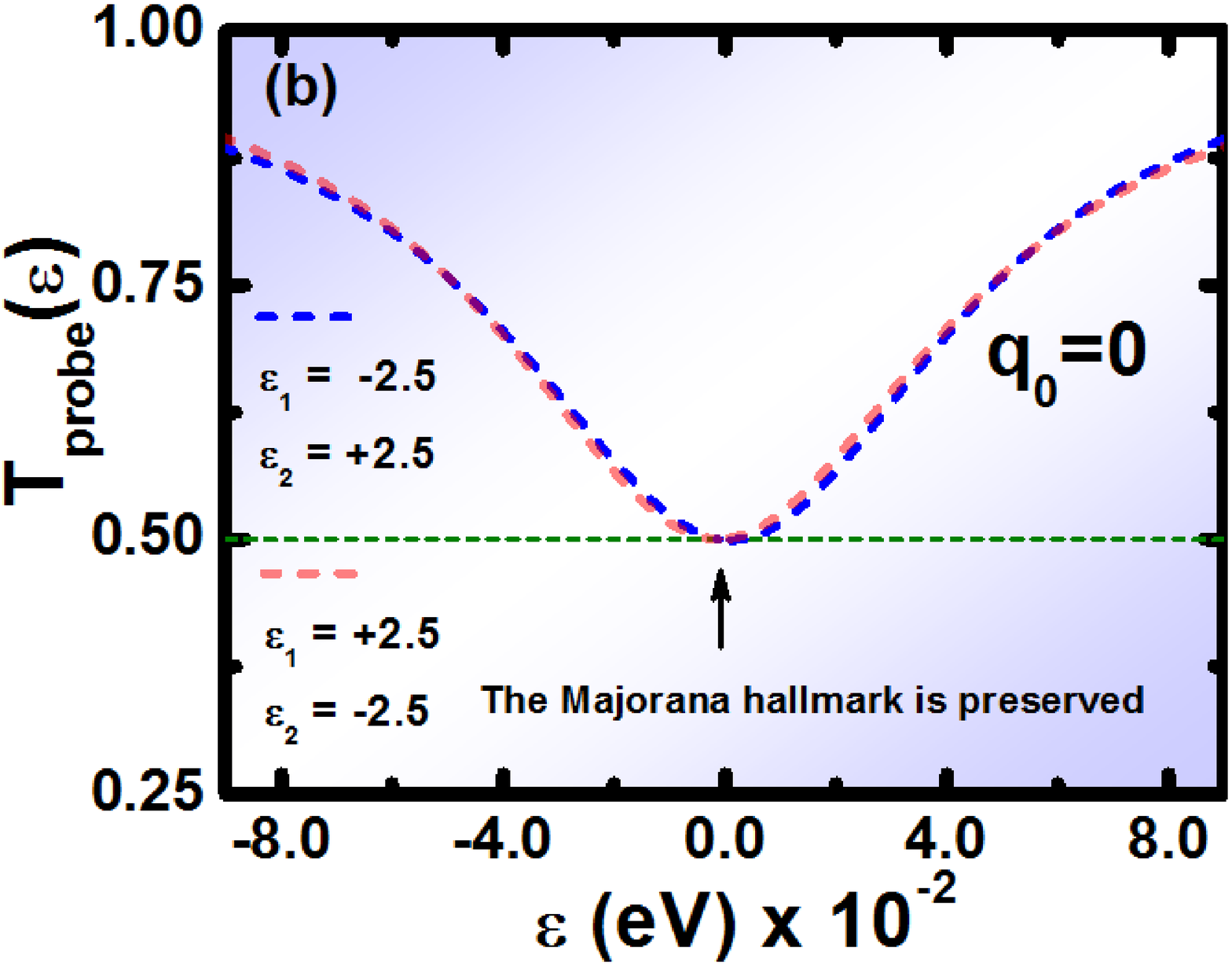}}}
\caption{(Color online) Parameters employed: $\epsilon_{M}=0$ {[}long enough
Kitaev chain{]}, $\lambda=5\Delta$ and $\Delta=\Gamma=0.2$ {[}see
Eqs. (\ref{eq:TIAM}), (\ref{eq:AP}) and (\ref{eq:AP_2}){]}. Transmittance
as a function of the single particle energy $\varepsilon$. Implementation
of the two-stage procedure of Sec. \ref{sec:intro}: (a) via the transmittance
$T_{\text{{full}}}(\varepsilon)$ of Eq. (\ref{eq:trans2}) in the
Fano regime $q_{b}=0$ {[}Eq. (\ref{eq:Fano_qb}){]}, where we see
the formation of a Majorana dip with an amplitude that fluctuates
slightly around $1/2$ (Majorana hallmark), but it remains pinned
at zero-bias even by performing the gates swap. In panel (b), the
transmittance $T_{\text{{probe}}}(\varepsilon)$ of Eq. (\ref{eq:LDOS_p1})
for $q_{0}=0$ does not exhibit such a fluctuation. The Majorana hallmark
remains unchanged for the STM tip considered as a probe.}

\label{Fig3}
\end{figure}

\begin{figure}[h]
\centerline{\resizebox{3.5in}{!}{\includegraphics[clip,width=0.14\textwidth,height=0.07\textheight]{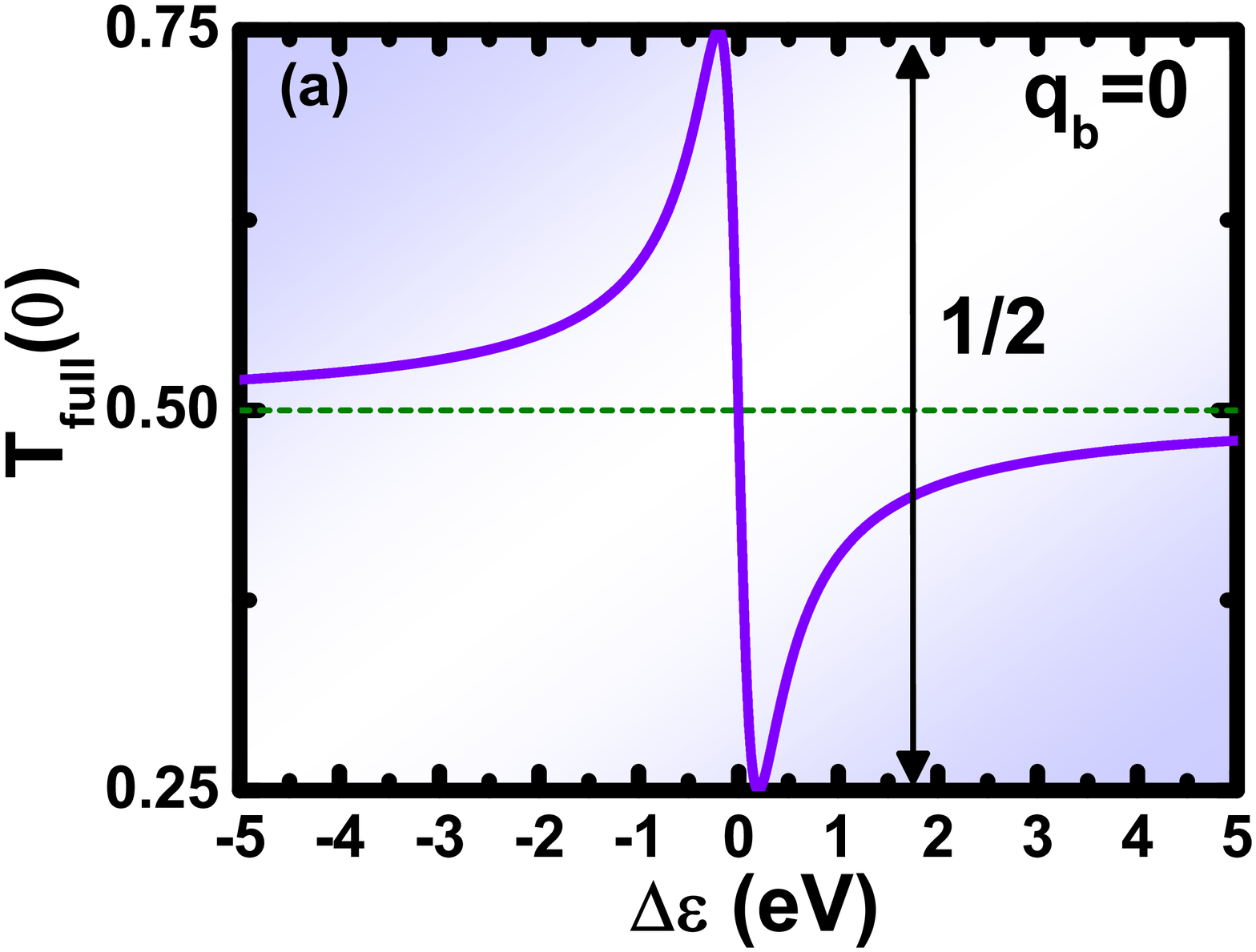}}}\centerline{\resizebox{3.5in}{!}{\includegraphics[clip,width=0.14\textwidth,height=0.07\textheight]{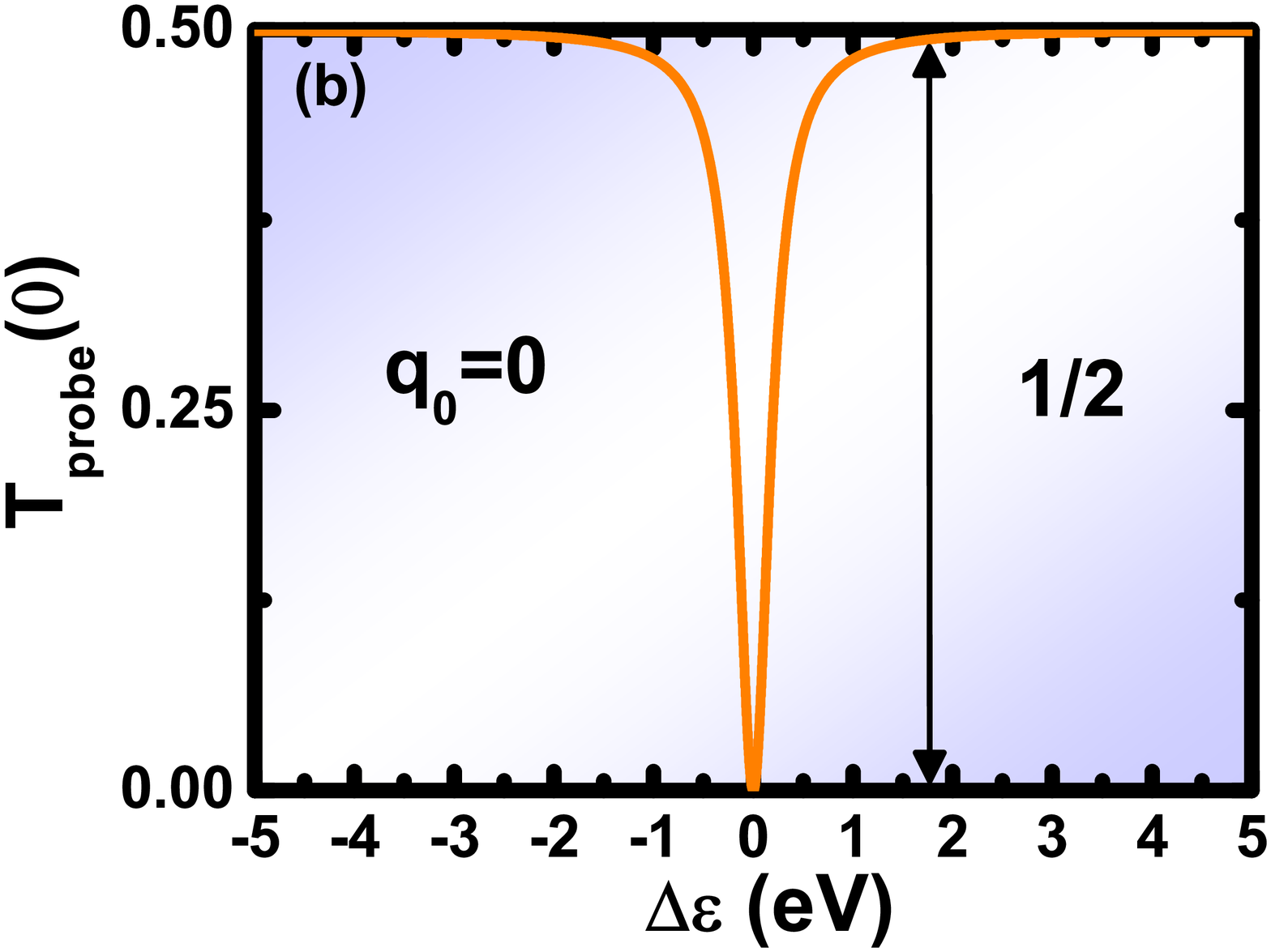}}}
\caption{(Color online) Parameters employed: $\epsilon_{M}=0$ {[}long enough
Kitaev chain{]}, $\lambda=5\Delta$ and $\Delta=\Gamma=0.2$ {[}see
Eqs. (\ref{eq:TIAM}), (\ref{eq:AP}) and (\ref{eq:AP_2}){]}. Panel
(a): Transmittance $T_{\text{{full}}}(0)$ of Eq. (\ref{eq:trans2})
in the Fano regime $q_{b}=0$ {[}Eq. (\ref{eq:Fano_qb}){]} as a function
of the symmetric detuning $\Delta\varepsilon=\varepsilon_{2}-\varepsilon_{1}$.
For the STM tip in the same footing as the ``host+adatoms'' system,
we see a novel feature in the transmittance profile: an unexpected
Fano lineshape emerges and the Fano dip is not verified. Pained (b):
in the case of the STM tip as a probe, the transmittance $T_{\text{{probe}}}(0)$
of Eq. (\ref{eq:LDOS_p1}) with $q_{0}=0$ leads to the standard Fano
antiresonance. We remark that despite the small difference in the
Majorana dip of Fig. \ref{Fig3}(a) with respect to that found in
Fig. \ref{Fig3}(b), the zero-bias transmittance as a function of
the detuning $\Delta\varepsilon$, yields two distinct lineshapes.
However, in both situations, the transmittance does not exceed an
amplitude of $1/2$.}

\label{Fig4}
\end{figure}

\begin{figure}[h]
\centerline{\resizebox{3.5in}{!}{\includegraphics[clip,width=0.14\textwidth,height=0.14\textheight]{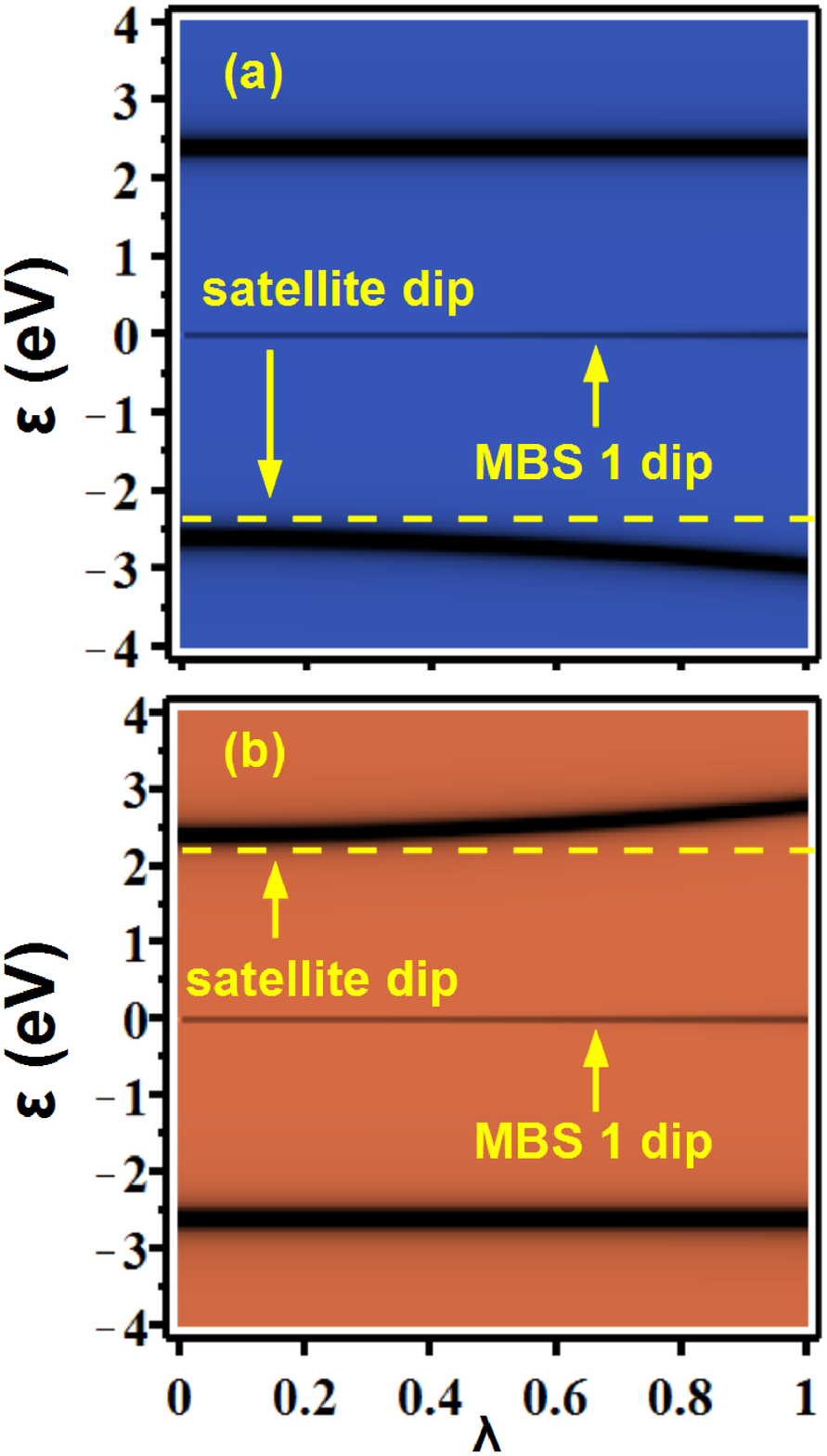}}}
\caption{(Color online) Parameters employed: $\epsilon_{M}=0$ {[}long enough
Kitaev chain{]}, $\Delta=0.2$ {[}see Eqs. (\ref{eq:TIAM}) and (\ref{eq:AP_2}){]}.
Density plots of the transmittance $T_{\text{{full}}}(\varepsilon)$
determined by Eq. (\ref{eq:trans2}) in the Fano regime $q_{b}=0$
{[}Eq. (\ref{eq:Fano_qb}){]} as a function of the single particle
energy $\varepsilon$ and the coupling $\lambda$ in units of $\Delta$.
Implementation of the two-stage procedure of Sec. \ref{sec:intro}:
(a) $\Delta\varepsilon=+5$ (b) $\Delta\varepsilon=-5$. Here we see
the main result of this procedure: the formation of a Majorana dip
(in black) pinned at zero-bias even by performing the gates swap.
The satellite dips (also in black) do not share such a feature being
significantly shifted under the permutation of the levels in the adatoms.}

\label{Fig5}
\end{figure}

In Fig. \ref{Fig2}(a) for the case of a free system, i.e., without
MBSs {[}solid-green curve{]}, we observe two antiresonances, each
one placed around the corresponding adatom level given by $\varepsilon_{1}=-2.5$
and $\varepsilon_{2}=+2.5$, respectively. We name these antiresonances
as satellite dips. Off the antiresonances, the transmittance approaches
the unitary limit and the conductance reaches $\mathcal{G}=\mathcal{G}_{0}=e^{2}/h$.
Notice, for instance, the central region bounded by the range $-1.75\lesssim e\varphi\lesssim1.5$
{[}shaded region{]}, where we have a ballistic plateau with the aforementioned
conductance. In this free system, the Green's functions of the model
are symmetric under the permutation of the indexes that designate
the parameters of the adatoms. This property is confirmed by the corresponding
solid-green curve of Fig. \ref{Fig2}(b), obtained with $\varepsilon_{1}=+2.5$
and $\varepsilon_{2}=-2.5$, which agrees with that for $\varepsilon_{1}=-2.5$
and $\varepsilon_{2}=+2.5$ in Fig. \ref{Fig2}(a). Therefore, the
two-stage procedure proposed in this work, in particular for the case
of a free double adatom system, yields two identical curves for the
transmittance. However, for the device side-coupled to the MBS 1,
a novel feature emerges in the central region.

By fixing $\varepsilon_{1}=-2.5$ and $\varepsilon_{2}=+2.5$, a dip
of amplitude nearby $1/2$ arises in the middle of the ballistic plateau,
due to the MBS 1 attached to the adatom 1 {[}see the dashed-blue curve
in Fig. \ref{Fig2}(a){]}. For this situation, the dip around $\varepsilon_{2}=+2.5$
coincides with the corresponding one found in the solid-green curve
of the free setup, which is due to the adatom 2 decoupled from the
MBS 1. Moreover, the antiresonance in the vicinity of $\varepsilon_{1}=-2.5$
is not coincident with that in Fig. \ref{Fig2}(a) of the solid-green
curve for the free system. As we can see, the position of such an
antiresonance is shifted as the aftermath of the coupling between
the MBS 1 and the adatom 1. After the swap procedure, which leads
to $\varepsilon_{1}=+2.5$ and $\varepsilon_{2}=-2.5$, the satellite
dips of Fig. \ref{Fig2}(b) {[}see the dashed-red lineshape{]} become
reversed with respect to those found in Fig. \ref{Fig2}(a).

We emphasize that the central antiresonance remains placed at zero-bias,
but its amplitude fluctuates slightly around $1/2$. This behavior
of the central dip appears in Fig. \ref{Fig3}(a), which can be clearly
visualized in the dashed-blue and red lineshapes, respectively. Therefore,
a pinned antiresonance protected against the two-stage procedure emerges
in the transmittance, which is placed at the zero-bias and characterized
by an amplitude nearby $1/2$. In contrast, the satellite dips do
not share such a pinning, since they move significantly under the
permutation of the levels in the adatoms. However, the complete robustness
of the Majorana hallmark does not exist anymore as found in Refs.
{[}\onlinecite{key-25}{]} and {[}\onlinecite{key-27}{]}: the
amplitude is not fixed at $1/2$ as a straight result of the interplay
between the adatom decoupled from the Kitaev chain and the Fano regime
as well, obtained with $x=1$ in Eq. (\ref{eq:Fano_qb}). In this
situation, the real and imaginary parts of the self-energy $\Sigma$,
which read $\Sigma^{R}$ and $\Sigma^{I},$ respectively given by
Eqs. (\ref{eq:setil}) and (\ref{eq:gtil}), depend on $x$. Otherwise,
it would correspond to the case of the tip considered as a probe of
the LDOS for the ``host+adatoms'' system, which suppresses the fluctuation
of the Majorana hallmark. This feature can be observed by using the
transmittance $T_{\text{{probe}}}(\varepsilon)$ of Eq. (\ref{eq:LDOS_p1})
with $q_{0}=0$, which is confirmed by the dashed-blue and red lineshapes
of Fig. \ref{Fig3}(b). In fact, it can be observed an antiresonance
pinned at zero-bias characterized by a constant amplitude of $1/2$.
In this case, $\Sigma^{R}$ and $\Sigma^{I}$ do not depend on $x$,
since $x\ll1$ for a probe tip {[}see Eq. (\ref{eq:Tb1}){]}. As a
result, the Majorana hallmark is preserved under the gates swap.

Thus in order to explore the effects due to the fluctuation of the
zero-bias transmittance, we present the analysis of $T_{\text{{full}}}(0)$
and $T_{\text{{probe}}}(0)$ as a function of the symmetric detuning
$\Delta\varepsilon$. In both cases, the Fano parameters are $q_{b}=0$
and $q_{0}=0$, which according to Fano's theory, lead to a destructive
interference pattern. Such a behavior can be seen in the transmittance
versus $\varepsilon$ plots of Figs. \ref{Fig3}(a) and (b). Additionally,
we point out that the Majorana dip verified in the former differs
slightly with respect to that found in the latter. Remarkably, the
slight fluctuation of the Majorana hallmark in Fig. \ref{Fig3}(a)
is able to provide an unexpected profile of $T_{\text{{full}}}(0)$
versus $\Delta\varepsilon$, which differs expressively of a Fano
dip. The result of this analysis appears in the solid-violet curve
of Fig. \ref{Fig4}(a), where it is observed that the transmittance
approaches 1/2 from upper (lower) values for $\Delta\varepsilon<0$
($\Delta\varepsilon>0$). In the domain of $\Delta\varepsilon<0$,
it reaches the maximum value of $3/4$, while for $\Delta\varepsilon>0$,
it decreases to $1/4$. Notice that the variation of the transmittance
with $\Delta\varepsilon$ does not exceed an amplitude of $1/2$ and
particularly at $\Delta\varepsilon=0$, the transmittance recovers
the Majorana hallmark $1/2$. On the other hand, in Fig. \ref{Fig4}(b),
the transmittance $T_{\text{{probe}}}(0)$ as a function of $\Delta\varepsilon$
in the solid-orange curve, displays the standard profile of Fano antiresonance
for $q_{0}=0$. Notice that in both Figs. \ref{Fig4}(a) and (b),
the variation of the transmittance with $\Delta\varepsilon$ is $1/2$.
We highlight that the unexpected Fano profile found in this work becomes
a way to identify the existence of isolated MBSs, since the lineshape
in Fig. \ref{Fig4}(a) is due to a long enough Kitaev chain within
the topological phase.

In summary, despite the same Fano parameters $q_{0}=0$ and $q_{b}=0$
in $T_{\text{{probe}}}(0)$ and $T_{\text{{full}}}(0)$, respectively
for Eqs. (\ref{eq:LDOS_p1}) and (\ref{eq:trans2}), which lead to
Fano dips slightly different as those found in Figs. \ref{Fig3}(a)
and (b), we demonstrate in this work that the usual hypothesis of
the STM tip acting as a probe is insensitive for the complete knowing
of the zero-bias transmittance versus the symmetric detuning $\Delta\varepsilon$.
To overcome such an obstacle, the proper description should consider
the STM tip in the same footing as the ``host+adatoms'' system.
It is worth mentioning that we do not present the results for the
case $q_{b}\gg1$, since it still obeys the standard Fano's theory,
which gives a resonance profile in the $T_{\text{{full}}}(0)$ versus
$\Delta\varepsilon$ plot as expected. In Fig. \ref{Fig5} the density
plots for $T_{\text{{full}}}(\varepsilon)$ of Eq. (\ref{eq:trans2})
with $q_{b}=0$ as a function of $\varepsilon$ and the coupling $\lambda$
are shown. In these graphs, dips appear (black color regions) being
possible to observe that the MBS 1 dip at zero-bias is the only structure
that does not change with the implementation of the two-stage procedure
as well as with the increase of $\lambda$. On the other hand, the
positions of the satellite dips are displaced by changing $\lambda$
and no pinning is observed. This feature can be visualized in the
dips that deviate from the yellow-dashed lines in Figs. \ref{Fig5}(a)
and (b), respectively for $\Delta\varepsilon=+5$ and $\Delta\varepsilon=-5$.

\section{Conclusions}

\label{sec4}

We have explored theoretically in the context of quantum transport
an effective Hamiltonian supporting Majorana quasiparticles for a
long enough Kitaev chain in the topological phase. This system is
coupled to a setup made by an STM tip and a metallic host with two
adatoms. Our analysis has revealed that the Green's functions of the
adatoms become symmetric by neglecting the hopping term between one
adatom and a side-coupled MBS. However, if we consider this parameter
relevant, a lack of symmetry manifests in these functions.

To read out this feature experimentally, it has been proposed a two-stage
procedure of gates swap by using AFM tips. As a result, a persistent
zero-bias dip with an amplitude nearby $1/2$ emerges in the transmittance
arising from the isolated MBS under the aforementioned procedure.
We have also verified that the fluctuation of the Majorana hallmark
occurs only for the STM tip treated in the same footing as the ``host+adatoms''
system. In the case of an STM tip as a probe, the robustness of the
Majorana hallmark is kept. However, this small difference between
these two Majorana dips results in contrasting Fano profiles for the
zero-bias transmittance versus the symmetric detuning. In the case
of the STM tip acting as probe, Fano's theory is confirmed, but with
the tip in the same footing as the ``host+adatoms'' system, an unexpected
Fano lineshape appears. We conclude that to access this non trivial
Fano profile, the assumption of an STM tip acting as a probe should
not be used.
\begin{acknowledgments}
The authors thank Drs. E. Vernek and J. C. Egues for valuable discussions.
This work was supported by the Brazilian agencies CNPq, CAPES and
PROPe/UNESP. \end{acknowledgments}

\end{document}